\documentclass[12pt,twoside]{article}
{\catcode `\@=11 \global\let\AddToReset=\@addtoreset}
\AddToReset{equation}{section}

\def\greaterthansquiggle{\raise.3ex\hbox{$>$\kern-.75em\lower1ex\hbox{$\sim$}}}
\def\lessthansquiggle{\raise.3ex\hbox{$<$\kern-.75em\lower1ex\hbox{$\sim$}}}

\newcommand{\beq}{\begin{equation}}
\newcommand{\eeq}{\end{equation}}
\newcommand{\beqa}{\begin{eqnarray}}
\newcommand{\eeqa}{\end{eqnarray}}
\newcommand{\beqan}{\begin{eqnarray*}}
\newcommand{\eeqan}{\end{eqnarray*}}
\newcommand{\ba}{\begin{array}}
\newcommand{\ea}{\end{array}}
\newcommand{\no}{\nonumber}

\newcommand{\ol}{\overline}
\newcommand{\ra}{\rightarrow}

\newcommand{\ve}{\varepsilon}

\newcommand{\V}{{\cal V}}

\def\nz{\ifmmode {I\hskip -3pt N} \else {\hbox {$I\hskip -3pt N$}}\fi}
\def\zz{\ifmmode {Z\hskip -4.8pt Z} \else
       {\hbox {$Z\hskip -4.8pt Z$}}\fi}
\def\qz{\ifmmode {Q\hskip -5.0pt\vrule height6.0pt depth 0pt
       \hskip 6pt} \else {\hbox
       {$Q\hskip -5.0pt\vrule height6.0pt depth 0pt\hskip 6pt$}}\fi}
\def\rz{\ifmmode {I\hskip -3pt R} \else {\hbox {$I\hskip -3pt R$}}\fi}
\def\cz{\ifmmode {C\hskip -4.8pt\vrule height5.8pt\hskip 6.3pt} \else
       {\hbox {$C\hskip -4.8pt\vrule height5.8pt\hskip 6.3pt$}}\fi}
\def\au{{\setbox0=\hbox{\lower1.36775ex%
\hbox{''}\kern-.05em}\dp0=.36775ex\hskip0pt\box0}}
\def\ao{{}\kern-.10em\hbox{``}}

\begin{document}
\begin{titlepage}
\begin{flushright} UWThPh-2000-2\\ 
January 10, 2000
\end{flushright}

\vspace*{2.5cm}
\begin{center}
{\Large \bf A mixed mean-field/BCS phase\\[10pt] with an energy
gap at high $T_c$ }\\[30pt] N. Ilieva$^{\ast,\sharp}$ and W.
Thirring
\\ [10pt] Institut f\"ur Theoretische Physik \\ Universit\"at
Wien\\
\smallskip
and \\
\smallskip
Erwin Schr\"odinger International Institute\\ for Mathematical
Physics\\

\vfill 
\vspace{0.8cm}

\begin{abstract}
We construct a pair potential which in a scaling limit leads to a
Hamiltonian that generates co-existing mean-field and
superconducting phases. Depending on the relative values of
the coupling constants, the superconducting phase may exist
at arbitrarily high temperatures.

\vfill \vspace{0.4cm} {\small PACS codes: 74.20.Fg, 05.30.Fk,
05.70.Fh

\vspace{0.2cm} Keywords: high-temperature superconductivity,
mean-field theory, pair potential, phase transitions}

\end{abstract}
\end{center}

\vfill {\footnotesize

$^\star$ Work supported in part by ``Fonds zur F\"orderung der
wissenschaftlichen Forschung in \"Osterreich" under grant
P11287--PHY;

$^\ast$ On leave from Institute for Nuclear Research and Nuclear
Energy, Bulgarian Academy of Sciences, Boul.Tzarigradsko Chaussee
72, 1784 Sofia, Bulgaria

$^\sharp$ E--mail address: ilieva@ap.univie.ac.at}
\end{titlepage}

\vfill \eject
\setcounter{page}{2}

\section*{Introduction}

In quantum mechanics a mean field theory means that the particle
density $\rho(x) = \psi^*(x)\psi(x)$ (in second quantization)
tends to a {\it c}-number in a suitable scaling limit. Of course,
$\rho(x)$ is only an operator valued distribution and the smeared
densities $\rho_f = \int dx\,\rho(x)f(x)$ are (at best) unbounded
operators, so norm convergence is not possible. The best one can
hope for is strong resolvent convergence in a representation where
the macroscopic density is built in. The BCS-theory of
superconductivity is of a different type where pairs of creation
operators with opposite momentum
$\tilde\psi^*(k)\,\tilde\psi^*(-k)$ ($\tilde\psi$ the Fourier
transform and with the same provisio) tend to {\it c}-numbers.
This requires different types of correlations and one might think
that the two possibilities are mutually exclusive. We shall show
that this is not so by constructing a pair potential where both
phenomena occure simultaneously. On purpose we shall use only one
type of fermions as one might think that the spin-up electrons
have one type of correlation and the spin-down the other. Also the
state which carries both correlations is not an artificial
construction but it is the KMS-state of the corresponding
Bogoliubov Hamiltonian. Whether the phenomenon occurs or not
depends on whether the emerging two coupled ``gap equations" have
a solution or not, which happens to be the case in certain regions
of the parameter space (temperature, chemical potential, relative
values of the two coupling constants). Moreover, in the new phases
with $\lambda_B, \lambda_M<0$ transition temperature $T_c$ may
become arbitrarily high. Our considerations hold for arbitrary
space dimension.

\section{Quadratic fluctuations in a KMS-state}

The solvability of the BCS-model \cite{BCS} rests upon the
observation \cite{H} that in an irreducible representation the
space average of a quasi-local quantity is a {\it c}-number and is
equal to its ground state expectation value. This allows one to
replace the model Hamiltonian by an equivalent approximating one
\cite{NNB}. Remember that two Hamiltonians are considered to be
equivalent when they lead to the same time evolution of the local
observables \cite{TW}.

The same property holds on also in a temperature state (the
KMS-state) and under conditions to be specified later it makes the
co-existence of other types of phases possible.

To make this apparent, consider the approximating (Bogoliubov)
Hamiltonian \beqa H_B' &=& \int dp\,\left\{\omega(p)a^*(p)a(p) +
\frac{1}{2}\Delta_B(p)\left[a^*(p)a^*(-p) +
a(-p)a(p)\right]\right\} \no \\ &=& \int W(p) b^*(p)b(p)\, , \eeqa
which has been diagonalized by means of a standard Bogoliubov
transformation with real coefficients (the irrelevant infinite
constant in $H_B'$ has been omitted) $$ b(p) = c(p)a(p) +
s(p)a^*(-p)\,, \qquad a(p) = c(p)b(p) - s(p)b^*(-p) $$ with \beq
c(p) = c(-p)\,, \qquad s(p) = -s(-p)\,,\qquad c^2(p) + s^2(p) =
1\, , \eeq so that the following relations hold (keeping in mind
that$\Delta, W, s, c$ will be $\beta$--dependent) $$ W(p) =
\sqrt{\omega^2(p) + \Delta_B^2(p)} = W(-p) $$ \beq c^2(p) - s^2(p)
= \omega(p)/W(p)\,, \qquad 2c(p)s(p) = \Delta_B(p)/W(p) \eeq

Hamiltonian (1.1) generates a well defined time evolution and a
KMS-state for the $b$-operators. For the original creation and
annihilation operators $a, a^*$ this gives the following evolution
$$ a(p) \ra a(p)\left(c^2(p)e^{-iW(p)t} + s^2(p)e^{iW(p)t}\right)
- 2ia^*(-p)c(p)s(p)\sin W(p)t $$ and nonvanishing termal
expectations \beqa \langle a^*(p)a(p')\rangle &=&
\delta(p-p')\left\{\frac{c^2(p)}{1+e^{\beta(W(p)-\mu)}} +
\frac{s^2(p)}{1+e^{-\beta(W(p)-\mu)}}\right\} \no \\[2pt] &:=&
\delta(p-p')\{p\}\\ \langle a(p)a(-p')\rangle &=&
\delta(p-p')c(p)s(p)\tanh\frac{\beta(W(p)-\mu)}{2} :=
\delta(p-p')[p] \eeqa $$ \{p\} = \{-p\}, \qquad [p] = -[-p] $$ $c$
and $s$ are multiplication operators and are never
Hilbert--Schmidt. Thus different $c$ and $s$ lead to inequivalent
representations and should be considered as different phases of
the system.

The expectation value of a biquadratic (in creation and annihilation
operators) quantity is expressed through (1.4,5)
\beqa
\langle a^*(q)a^*(q')a(p)a(p')\rangle =
\delta(q+q')\delta(p+p')[q][p]-& \no \\
- \delta(p-q)\delta(p'-q')\{p\}\{p'\} +
\delta(p-q')\delta(p'-q)\{p\}\{p'\} &
\eeqa

So far we have written everything in terms of the operator valued
distributions $a(p)$. They can be easily converted into operators
in the Hilbert space generated by the KMS-state by smearing with
suitable test functions. Thus, by smearing with e.g. \beq
e^{-\kappa (p+p')^2-\kappa (q+q')^2}v(p)v(q), \qquad v \in
L_2({\bf R}^d) \eeq one observes that in the limit $\kappa \ra
\infty$ the first term in (1.6) remains finite $$ 0 < \int
dpdqv(p)v(q)[p][q] < \infty\, , $$ while the two others vanish $$
\lim_{\kappa\ra\infty} \int dpdp'
e^{-2\kappa(p+p')^2}v(p)v(p')\{p\}\{p'\} = \lim_{\kappa\ra\infty}
\kappa^{-3/2}\int dpv^2(p)\{p\}^2 = 0. $$

Since we are in the situation of {\it Lemma 1} in \cite{IT}, we have thus
proved the following statement
\beq
\mbox{s-}\lim_{\kappa\ra\infty}\int
dpdp'\V(q,q',p,p')e^{-\kappa(p+p')^2}a(p)a(p') = \int
dp\V(q,q',p,-p)[p]
\eeq
for kernels $\V$ such that the integrals are finite.

With this observation in mind, a potential which acts for $\kappa
\ra \infty$ like (1.1) might be written as \beq V_B = \kappa^{3/2}
\int dpdp'dqdq'\,a^*(q)a^*(q')a(p)a(p')\V_B(q,q',p,p')\,
e^{-\kappa(p+p')^2-\kappa(q+q')^2} \eeq with $\V_B(q,q',p,p') =
-\V_B(q',q,p,p')$ etc., in order to respect the Fermi-nature of
$a$'s. This potential has the property $$
\begin{array}{ll}
\Vert V \Vert < \infty & \qquad \mbox{ for }\, \kappa <\infty \no \\[4pt]
\Vert V \Vert \ra \infty & \qquad \mbox{ for }\, \kappa \ra \infty \no
\end{array}
$$ Despite this divergence, potential (1.9) may still generate a
well-defined time evolution. The strong resolvent convergence in
(1.8) is essential, weak convergence would not be enough since it
does not guarantee the automorphism property $$ \tau_\kappa^t(ab)
= \tau_\kappa^t(a)\tau_\kappa^t(b)\,\ra\, \tau_\infty^t(ab) =
\tau_\infty^t(a)\tau_\infty^t(b)\,. $$ Note that the parameter
$\kappa$ plays in this construction the role of the volume from
the considerations in \cite{H}.

In the mean-field regime we want an effective Hamiltonian \beq
H_B'' = \int dp \left[ \omega(p)a^*(p)a(p) +
\Delta_M(p)a^*(p)a(p)\right]\, . \eeq Here the KMS-state is
defined for the operators $a, a^*$ themselves and one should
rather smear by means of \beq
e^{-\kappa(q-p)^2-\kappa(q'-p')^2}v(p)v(p') \eeq instead of (1.7),
thus concluding that \beq \mbox{ s-}\lim_{\kappa\ra\infty} \int
dpdq e^{-\kappa(q-p)^2} a^*(q)a(p) \V_M(q,q',p,p') = -\int
dp\frac{\V_M(p,q',p,p')}{1 + e^{\beta(\ve(p)-\mu)}}\, , \eeq with
$\ve(p) = \omega(p) + \Delta_M(p)$. Relation (1.12) then suggests
another starting potential \beq V_M = \kappa^{3/2} \int
dpdp'dqdq'\,a^*(q)a^*(q')a(p)a(p')\V_M(q,q',p,p')\,
e^{-\kappa(q-p)^2-\kappa(q'-p')^2} \eeq with the same symmetry for
the density $\V_M$ as in (1.9). However, in both cases a Gaussian
form factor in the smearing functions (1.7),(1.11) has been chosen
just for simplicity. In principle, this might be $C_o^\infty$
functions which have the $\delta$-function as a limit.

\section{The model}

Consider the following Hamiltonian \beq H = H_{kin} +  V_B +  V_M
\, , \eeq where $H_{kin}$ is the kinetic term and $ V_B, V_M$ are
given by (1.9),(1.13). The solvability of the model for $\kappa
\ra \infty$ depends on whether or not it would be possible to
replace (2.1) by an equivalent Hamiltonian that might be readily
diagonalized.  The object of interest is the commutator of, say, a
creation operator with the potential. With (1.8), (1.12) taken
into account, it reads \beq [a(k), V] = 2\int
dp\left\{c(p)s(p)\,[p]\,\V_B(k,-k,p,-p)a^*(-k) +
\V_M(p,k,p,k)\,\{p\}\,a(k)\right\} \eeq The Bogoliubov-type
Hamiltonian for our problem should be a combination of (1.1) and
(1.10), that is of the form \beq H_B = \int dp \left\{
a^*(p)a(p)[\omega(p) + \Delta_M(p)] +
\frac{1}{2}\Delta_B(p)[a^*(p)a^*(-p) + a(-p)a(p)]\right\} \eeq
This Hamiltonian becomes equivalent to the model Hamiltonian
(2.1), provided the commutator $[a(k), H_B - H_{kin}]$ equals
(2.2). Thus we are led to a system of two coupled ``gap equations"
\beqa \frac{1}{2} \Delta_M(k) &=& \int \V_M(k,p)\,
\left\{\frac{c^2(p)}{1 + e^{\beta(\ol W(p)-\mu)}} +
\frac{s^2(p)}{1+e^{-\beta(\ol W(p)-\mu)}}\right\} \,dp \\[10pt]
\Delta_B(k) &=& \int \V_B(k,p)\, \frac{\Delta_B(p)}{\ol W(p)}
\,\tanh\frac{\beta(\ol W(p)-\mu)}{2}\,dp\, , \eeqa with \beq \ol
W(p) = \sqrt{[\omega(p) + \Delta_M(p)]^2 + \Delta_B^2(p)}\, . \eeq
$c$ (and thus $s$, Eq.(1.2)) are determined by either of the
following conditions \beq c^2(p) - s^2(p) =
[\omega(p)+\Delta_M(p)]/\ol W(p)\,, \qquad 2c(p)s(p) =
\Delta_B(p)/\ol W(p)\, . \eeq The temperature and the
interaction-strenght dependence of the system (2.4--7) encode the
solvability of the model \cite{NPB}.

\section{High $T_c$ case}
We are now looking for a mechanism for high temperature
superconductivity, i.e. a high $T_c$ where $\Delta_B$ starts to
vanish. If we make the ansatz $$ \V_B(k,p) = \lambda_Bv(k)v(p)\,,
\qquad \int v^2(p)dp = 1\,, \qquad v(p) = -v(-p)\,, $$ then
(2.5)becomes $$ \Delta_B(k) = \lambda_Bv(k)\int
dp\frac{v(p)\Delta_B(p)}{\ol W(p)}\tanh{\frac{\beta(\ol
W(p)-\mu)}{2}\, .} $$ For $\lambda_B<0$ we must have $\ol W<\mu$
and since $\tanh x < x \,,\forall x>0$, we conclude that $$
T<\frac{|\lambda_B|}{2}\int dp v^2(p)\left(\frac{\mu}
{\ol W(p)}-1\right)\, . $$ If $\Delta_B$ starts to vanish, $\ol W(p) =
|\omega(p)+\Delta_M(p)|$, so if $\Delta_M<0$ and near $\omega(p)$,
$T_c$ can become arbitrarily high $$ T_c <
\frac{|\lambda_B|}{2}\left(-1+\mu\int\frac{dp
v^2(p)}{|\omega(p)+\Delta_M(p)|}\right)\, . $$ Thus a negative
mean field which almost cancels the kinetic energy $\omega$ gives
the electrons so much mobility to respond to $\lambda_B<0$ that
even at high temperatures a gap $\Delta_B$ can develope. There is
a small problem since $\Delta_B(-k)=-\Delta_B(k)$. However $v(k)$
need not be continuous and since only $\Delta_B^2$ enters in $\ol
W$ the gap parameter $\Delta_B^2(0)$ can effectively be $\not= 0$.
This problem disappears if we include spin and thus have
$a_\uparrow(p)a_\downarrow(-p)$ in $V_B$.

\section{Conclusion}
Our model has four parameters, $\,\lambda_M, \lambda_B, \mu, T\,$,
but by scaling only their ratios are essential. For infinite
temperature $\beta = 0\,$  Eqs.(3.1--3) admit only the mean field
solution $\Delta_B = 0\,,\, \Delta_M = \lambda_M,\,\, \ol W = \mu
+ \lambda_M$. By lowering the temperature one meets also the
BCS-type solution but in a rather complicated  region in the
3--dimensional parameter space.

Whenever $\lambda_B$ is positive, it must be also $ \,>\mu$. Also
for negative $\,\lambda_B,\, \lambda_M\,$ and $\,\lambda_M >
-\mu\,$ there exists a finite gap for $\lambda_B$. A perturbation
theory with respect to $\lambda_B$ is in general doomed to failure
since for no point on the $\lambda_B = 0$ axis there is a
neighbourhood full of the $\Delta_B \not= 0$ phase.

It is interesting that without a mean field (the $\lambda_M = 0$ axis)
there are superconducting solutions only for $\lambda_B > \mu$. An
attractive mean field ($\lambda_M < 0$) stimulates superconductivity since
then it also appears for negative $\lambda_B$. However, too strong mean
field attraction destroyes it again.

The most remarkable fact is that whilst for $\lambda>0$ the
temperature for a superconducting phase is limited as in the BCS
theory by $T\ll(\lambda_B-\mu)/2$, in the new phases for
$\lambda_B<0$, $\lambda_M<0$ we only get
$T<|\lambda_B||\lambda_M|/2(\mu-|\lambda_M|)$ and thus for
$\lambda_M \rightarrow -\mu$, $T$ can become arbitrarily big.

\section*{Acknowledgements}
We are grateful to D.Ya. Petrina for stimulating our interest into
the problem and to J. Brankov and N. Tonchev who shared with us
their experience in it (see, e.g. \cite{TB} and references
therein). We also appreciate suggestive discussions with H.
Narnhofer.

N.I. thanks the International Erwin Schr\"odinger Institute for
Mathematical Physics where the research has been performed, for
hospitality and financial support. This work has been supported
in part also by ``Fonds zur F\"orderung der wissenschaftlichen
Forschung in \"Osterreich" under grant P11287--PHY.

{}
\end{document}